\documentclass[twocolumn,prb,showpacs]{revtex4}

\usepackage{graphicx}
\begin{document}
\draft
\title{Valence instability of cerium under pressure in the Kondo-like perovskite La$_{0.1}$Ce$_{0.4}$Sr$_{0.5}$MnO$_3$}
\author{T. Eto}
\address{Kumamoto Technology Center, Sony Semiconductor Kyushu Company Limited, Haramizu 4000-1,
Kikuyou, Kikuchi, Kumamoto 869-1102, Japan}
\author{A. Sundaresan}
\email{sundaresan@jncasr.ac.in}
\address{Chemistry and Physics of Materials Unit, Jawaharlal Nehru Centre for
 Advanced Scientific Research, Jakkur P. O. Bangalore 560 064 India}
\author{F. Honda}
\address{Advanced Science Research Center, Japan Atomic Energy
Research Institute, Tokai, Naka, Ibaraki 319-1195, Japan}
\author{G. Oomi}
\address{Department of Physics, Kyushu University, Ropponmatsu 4-2-1,
Fukuoka 810-8560, Japan}
\date{\today}
\begin{abstract}
Effect of  hydrostatic pressure and magnetic field on electrical
resistance of the Kondo-like perovskite manganese oxide,
La$_{0.1}$Ce$_{0.4}$Sr$_{0.5}$MnO$_3$ with a ferrimagnetic ground
state, have been investigated up to 2.1 GPa and  9 T. In this
compound, the Mn-moments undergo double exchange mediated
ferromagnetic ordering at $T_{\rm C}$ $\sim$ 280 K and there is a
resistance maximum, $T_{\rm max}$ at about 130 K which is
correlated with an antiferromagnetic ordering of {\it cerium} with
respect to the Mn-sublattice moments. Under pressure, the $T_{\rm
max}$ shifts to lower temperature at a rate of d$T_{max}$/d$P$ =
-162 K/GPa and disappears at a critical pressure $P_{\rm c}$
$\sim$ 0.9 GPa. Further, the coefficient, $m$ of $-logT$ term due
to Kondo scattering decreases linearly with increase of pressure
showing an inflection point in the vicinity of $P_{\rm c}$. These
results suggest that {\it cerium} undergoes a transition from
Ce$^{3+}$ state to Ce$^{4+}$/Ce$^{3+}$ mixed valence state under
pressure. In contrast to pressure effect, the applied magnetic
field shifts $T_{\rm max}$ to higher temperature presumably due to
enhanced ferromagnetic Mn moments.

\end{abstract}
\maketitle

Perovskite manganites $R_{1-x}A_{x}$MnO$_3$ ($R$ = trivalent rare
earth: $A$ = divalent alkaline earth) have been investigated
extensively due to a lot of fascinating electronic properties such
as insulator-metal transition (I-M), colossal magnetoresistance
(CMR) and charge, orbital and spin ordering. Magnetic field,
pressure, photons and electric field are the external parameters
by which the size, existence and interplay of these effects can be
varied.\cite{rao,Ramirez97} It is well known that the ground state
properties of half-doped manganites, $R_{0.5}A_{0.5}$MnO$_3$
depend upon the width ($W$) of one electron conduction band; $W$
is controlled by the average size ($r_A$) of the different
$A$-cations through Mn-O-Mn bond angle. For example,
La$_{0.5}$Sr$_{0.5}$MnO$_3$ with larger W ($r_A$ = 1.263 {\AA}) is
a metallic ferromagnet in the ground state.\cite{Jonker50} This
behavior has been explained by a theoretical scenario based on the
double exchange (DE) interaction between Mn$^{\rm 3+}$ and
Mn$^{\rm 4+}$ ions, where the spins of $e_{\rm g}$ electrons or
holes are aligned in the direction of the $t_{\rm 2g}$ local spins
by Hund coupling and electrical conduction arises due to the
hopping of the charge carriers from one ion to the next without
changing their spin orientation.\cite{Zener51,Anderson55,Gennes55}

\begin{figure}[b]
\includegraphics[width=7cm]{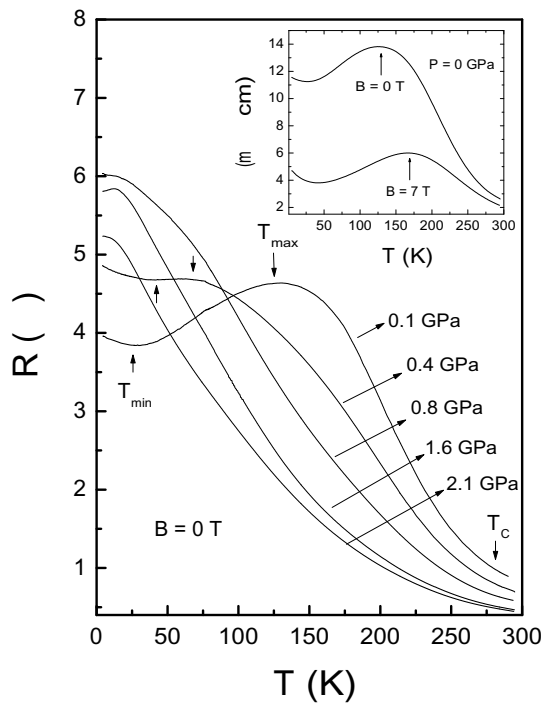}
 \caption[resist]{Temperature dependence of the electrical resistance $R$($T$)
of La$_{0.1}$Ce$_{0.4}$Sr$_{0.5}$MnO$_3$ at high pressures. The
downward and upward arrows indicate $T_{\rm max}$ and $T_{\rm
min}$, respectively. $T_{\rm C}$ is nearly 280 K. Inset shows the
shift of $T_{\rm max}$ to a higher temperature under a magnetic
field of 7 T and P = 0GPa.}
 \label{fig:resist}
\end{figure}
In compounds containing rare earth elements, {\it cerium} is well
known to exhibit a wide variety of electronic properties such as
valence change, heavy fermion, superconductivity and magnetic
ordering. In Pr$_{0.1}$Ce$_{0.4}$Sr$_{0.5}$MnO$_3$ compound having
larger $W$, Kondo like behavior and anomalous magnetic ordering of
Ce were reported.\cite{Sundaresan99} In this system, there is a
structural phase transition at 250 K  from the high temperature
orthorhombic phase (space group: $Imma$) to the low temperature
tetragonal phase ($I4/mcm$) which is accompanied by a
ferromagnetic ordering of manganese moments due to DE
interactions, and Ce moments order antiferromagnetically with
respect to Mn moments below T $\sim$ 120 K. The electrical
resistivity increases anomalously with decrease of temperature,
particularly below the Curie temperature $T_{\rm C}$, exhibiting a
resistivity maximum at 120 K ($T_{\rm max}$), which corresponds to
the ordering of Ce moments, and a minimum at 15 K ($T_{min}$). The
anomalous temperature dependence of resistivity $R(T)$ is in
contrast to the expected metallic behavior below $T_{\rm C}$ due
to DE interactions. Similar results were reported for
La$_{0.1}$Ce$_{0.4}$Sr$_{0.5}$MnO$_3$ with the ferromagnetic
ordering of Mn-moments at $T_{\rm C}$ $\sim$ 280 K and Ce ordering
below $\sim$ 130 K.\cite{Sundaresan01} In these systems with
larger $W$, since there is no charge ordering of manganese ions
and  the Mn-sublattice remains ferromagnetic down to 1.8 K, the
resistivity anomaly has been attributed to Kondo-like scattering
of Mn:$e_{\it g}$ conduction electrons by the localized Ce:4$f$
moments. The decrease of resistivity below $T_{max}$ or {\it
cerium} ordering temperature is due to reduced spin dependent
scattering.

In order to get more insight into the nature of interactions
between the localized Ce:4$f$ and ferromagnetic Mn moments, we
have investigated a simultaneous and individual effects of
hydrostatic pressure and magnetic field on the behavior of
electrical resistance in La$_{0.1}$Ce$_{0.4}$Sr$_{0.5}$MnO$_3$
within a temperature range of 4.2 K -- 300 K. We use this approach
because our earlier ambient-pressure work clearly established that
the resistivity features are correlated with magnetization and
neutron diffraction data.\cite{Sundaresan99,Sundaresan01}  From
this study, we substantiate that the anomalous temperature
dependence of electrical resistance is due to Kondo-like
scattering of Mn:$e_{\rm g}$ electrons by localized Ce:4$f$
moments. Further, we infer that there is a valence change of {\it
cerium} under pressure from Ce$^{3+}$ state to Ce$^{4+}$/Ce$^{3+}$
mixed valence state.


Polycrystalline sample of La$_{0.1}$Ce$_{0.4}$Sr$_{0.5}$MnO$_3$
was prepared by calcining stoichiometric mixtures of La$_2$O$_3$,
CeO$_2$, SrCO$_3$, Mn$_2$O$_3$ at 1100$^{\circ}$C and sintering at
1500$^{\circ}$C. Phase purity, nuclear and magnetic structures
were determined by powder x-ray and neutron diffraction
methods.\cite{Sundaresan01} The electrical resistance was measured
by a standard four-probe method with a direct current of 1 mA down
to 4.2 K. The electrical contact was made with a silver paint of
heat treatment type. Hydrostatic pressure up to 2.1 GPa was
generated by using a Cu-Be piston-cylinder device and 1:1 mixture
of Fluorinert FC70 and FC77 as a pressure medium. The pressure was
changed at RT to minimize internal strain in the specimen and the
load was controlled to within $\pm 1$ \% throughout the
measurement. The magnetic field was applied by a superconducting
magnet up to 9 T. The details of the present apparatus at
multi-extreme conditions have been reported previously.
\cite{Oomi97}
\begin{figure}[h]
\includegraphics[width=7cm]{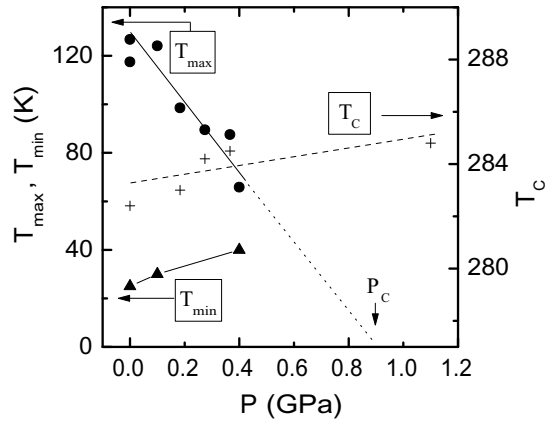}
 \caption[incline]{Pressure dependence of $T_{\rm max}$ and $T_{\rm min}$
at B = 0 T as shown by solid circle and triangle, respectively.
The solid, dashed and dotted lines are guides for the eye. $P_{\rm
c}$ is the value of pressure corresponds to the extrapolation of
$T_{max}$ to zero temperature.}
 \label{fig:temp}
\end{figure}

\begin{figure}[h]
\includegraphics[width=7cm]{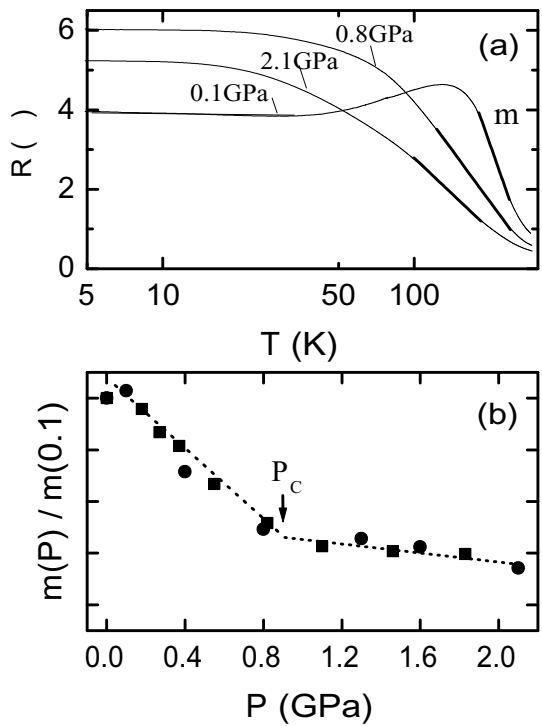}
 \caption[incline]{(a) Temperature dependence of $R$, in which the abscissa
is selected as a logarithmic scale and $-\log T$ dependence is
shown by solid lines. (b) Pressure dependence of the normalized
coefficient $m$($P$)/$m$(0.1) of $-\log T$ term. The previous data
of $m$($P$)/$m$(0.1) \cite{Eto01} is described as closed circle.
$P_{\rm c}$ is defined as an inflection point of the coefficient.
The dotted and solid lines are guides for the eye. }
 \label{fig:incline}
\end{figure}

\label{sec:results} Temperature dependence of electrical
resistance at various pressures is shown in Fig. \ref{fig:resist}.
In this figure, there is no indication for the DE ferromagnetic
metallic transition near $T_{\rm C}$ $\sim$ 280 K. However, we
have already reported\cite{Eto01} from the temperature derivative
of $R$ that, in this system, the $T_{\rm C}$ increases with
pressure at a rate of d$T_{\rm C}$/d$P$ = + 1.9 K/GPa, as shown in
Fig.\ref{fig:temp}. The small increase of $T_{\rm C}$ is
consistent with the fact that this system is in a weak coupling
region due to a large filling of conduction band, where the system
is less sensitive to pressure or change in W.\cite{ymori} At 0.1
GPa, $R$ increases logarithmically ($-\log T$) with decreasing
temperature in the range $\sim$ 180 -- 230 K as shown by solid
line in Fig. \ref{fig:incline}(a) and exhibits a broad maximum at
about 130 K ($T_{max}$) and a minimum at about 30 K ($T_{min}$)
which are shown by arrows in Fig.\ref{fig:resist}. As discussed
earlier,\cite{Sundaresan99} the $T_{max}$ is due to the onset of
antiferromagnetic ordering of Ce$^{3+}$ moments with respect to
the ferromagnetic Mn moments. This $R(T)$ behavior is reminiscent
of the typical concentrated Kondo
compounds.\cite{Kagayama91,Kagayama96} However, the origin of
$T_{max}$ in the present case should be compared and contrasted
with the concentrated Kondo systems where the origin of
resistivity maximum is due to crystal field effects and Kondo
coherence, and the $T_{max}$ has a large positive pressure
coefficient.\cite{Kagayama96,Oomi99} Further, the decrease of
resistivity below Ce ordering is not sharp such as one normally
observes at a magnetic ordering of Ce in intermetallics. This is
because, the antiferromagnetic coupling in the present system is
between Ce and Mn moments and the antiparallel arrangement of Ce
is induced by Mn moments.\cite{Sundaresan01} In fact, a recent
theoretical analysis on this system reproduce the magnetic
behavior of Mn and Ce in this system.\cite{avignon01}

In the present case, with increasing pressure, $T_{max}$ shifts to
a lower temperature at a rate of d$T_{max}$/d$P$ = -162 K /GPa,
which is in sharp contrast to the typical concentrated Kondo
systems, for example, CeAl$_3$ or CeInCu$_2$ where the $T_{max}$
increases with pressure.\cite{Kagayama91,Kagayama96} On the other
hand, $T_{min}$ seems to shift to higher temperature as shown in
Fig. \ref{fig:temp}. At pressures higher than 0.8 GPa both
$T_{max}$ and $T_{min}$ are not seen down to 4.2 K. It should be
noticed from the Fig. \ref{fig:temp} that the extrapolated value
of pressure, where the $T_{max}$ equals to zero, is 0.9 GPa, which
we termed as critical pressure $P_{\rm c}$. Since the $T_{max}$
corresponds to the ordering temperature of {\it cerium}, the
$P_{\it c}$ might indicate the suppression of magnetic ordering of
{\it cerium}. According to the effect of hydrostatic pressure on
DE ferromagnetism,\cite{ymori} we would expect that the $T_{max}$
associated with Ce ordering should increase with pressure. The
observed decrease of $T_{max}$ or the magnetic ordering
temperature of {\it cerium} with increase of pressure rather
indicates a valence change of {\it cerium} from Ce$^{3+}$ towards
Ce$^{4+}$ state. In contrast to pressure effect, the application
of magnetic field (B) shifts the $T_{max}$ to higher temperature
due to enhanced ferromagnetic Mn moments, as shown in the inset of
Fig.\ref{fig:resist}, for an applied field of 7 T. The presence of
resistivity minimum in the vicinity of 25 K in manganites is well
known and has been attributed to the effect of electron-electron
interactions.\cite{kumar}

It is intriguing and important to note the change of $R(T)$ at
various pressures as a function of temperature (see Fig.
\ref{fig:resist}). Near $T_{\rm C}$, the decrease of $R$ with
increase of pressure is due to; (i) widening of $e_{\rm g}$ band
as reported for La$_{1-x}$Sr$_x$MnO$_3$ \cite{ymori} and (ii)
valence change of {\it cerium} ion. On the other hand, at the
lowest temperature measured (4.2 K), $R$ increases with pressure
up to 0.8 GPa, close to $P_{\rm c}$, and then decreases with
further increase of pressure. Since the pressure effect on the DE
ferromagnet with larger $W$ does not have any contribution to $R$
at low temperatures, the increase of R suggests that the strength
of $J$, the exchange integral, which depends on the hybridization
between the conduction electron and $f$-electron, increases with
pressure. The decrease of $R$ above $P_{\it c}$ may be  due to
decrease of $T_{max}$. In the intermediate region (90 $\leq$ T
$\leq$ 250) the large decrease of $R$ with increase of pressure is
mainly due to suppression of spin dependent scattering that
results from the valence change of {\it cerium}. The fact that the
$R(T)$ does not become a metallic like even at 2.1 GPa suggests
that the {\it cerium} has not completely changed to Ce$^{4+}$
state and rather exists in Ce$^{4+}$/Ce$^{3+}$ mixed valence
state.
\begin{figure}[b]
 \includegraphics[width=7cm]{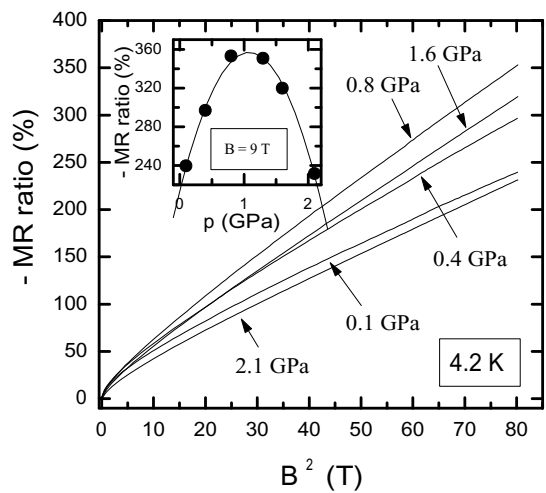}
 \caption[incline]{Pressure dependence of negative MR ratio
 (=$\Delta R/R$($B$)$\times$ 100 \%) at 4.2 K. The MR ratio at 9 T is plotted
in the inset as a function of pressure. A solid curve is guide for
the eye. }
 \label{fig:magresi}
\end{figure}
Fig.\ref{fig:incline}a, shows variation of $R$ at various
pressures as a function of temperature in logarithmic scale. It
can be seen that there is a region of $-logT$ dependence of $R$ at
high temperatures ($>$ 100 K) as shown by solid line at all
pressures. This is a signature of Kondo scattering of conduction
electrons by the localized moments. The coefficients of $-logT$
term is termed as {\it m}. In Fig.\ref{fig:incline}b, normalized
coefficient $m$($P$)/$m$(0.1) is plotted as a function of
pressure, where $m$($P$) is defined as $m$=$-\partial R
/\partial\log T$, and $m$(0.1) is that at 0.1 GPa. The
coefficient, $m$(P)/$m$(0.1) decreases linearly with increase of
pressure because of the change in valence state of {\it cerium} as
discussed above.  This behavior is also different qualitatively
from that of the typical Kondo compounds.\cite{Oomi99} Near
$P_{\rm c}$ $\sim$ 0.9 GPa, a small inflection is seen and above
which the effect of pressure on $m$ becomes smaller, indicating a
decrease in the strength of the Kondo scattering above $P_{\rm
c}$. This is consistent with the observation that the $T_{\rm
max}$ disappears at $P_{\rm c}$, where one would expect a change
in the interaction between the Ce:4$f$ local and the conduction
electron moments.

\begin{figure}[t]
\includegraphics[width=7cm]{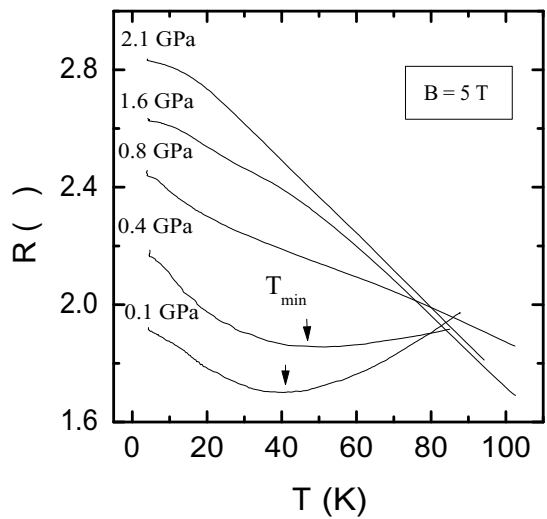}
 \caption[incline]{Pressure dependence of $R(T)$ at magnetic field of 5 T.
It can be seen that R at 4.2 K increases monotonically with
pressure contrary to zero applied field (see Fig.
\ref{fig:resist}). }
 \label{fig:resist5T}
\end{figure}
Magnetoresistance (MR) ratio ($\Delta
R/R$($B$)=($R$(0)-$R$($B$))/$R$($B$)) at various pressures at 4.2
K is plotted as a function of $B^2$ in Fig. \ref{fig:magresi}. The
negative CMR effect in perovskite manganese oxides is well known
as reported in previous experiments\cite{Ramirez97}. In the
present experiment, the negative MR monotonically increases with
applying $B$ which is mainly due to suppression of Kondo-like
scattering  and a small contribution from grain boundary
scattering, as the DE interactions in large $W$ materials results
in MR effect only near $T_{\rm C}$.\cite{Sundaresan99} Above 5 T,
MR lineary increases with respect to $B^2$, revealing the
existence of spin-dependent scattering. The inset shows MR ratio
at 4.2 K and at a field of 9 T as a function of pressure: MR
increases with pressure and a peak is clearly seen near $P_{\rm
c}$ $\sim$ 0.9 GPa. This is consistent with the observation that
the pressure dependence of $R$ at 4.2 K has similar behavior,
indicating the electronic changes associated with the valence
change of {\it cerium}.

The pressure dependence of $R(T)$ ( T $<$ 100 K) under a magnetic
field of $B$ = 5 T is shown in Fig. \ref{fig:resist5T}. By
applying 5 T, $T_{ min}$ shifts from 30 K to 40 K at 0.1 GPa. This
phenomenon was explained by the enhancement of {\it cerium}
ordering induced by the ferromagnetic manganese moments
\cite{Sundaresan99}. In contrast to zero applied field, $R$ at 4.2
K and 5 T increases with increasing pressure without showing any
maximum in the present pressure range. It is possible that $P_{\rm
c}$ exceeds the maximum pressure limit in our experiment because
the ordering temperature of {\it cerium} increases with applied
magnetic field.\cite{Sundaresan99} $T_{\rm min}$ increases with
pressure similar to that observed under zero field as shown Fig.
\ref{fig:temp}. It would be interesting to carry out neutron
diffraction experiment under pressure to see the changes in the
ordering temperature of {\it cerium}.

\label{sec:conclusion} In conclusion, we have found that under
zero applied magnetic field, the $T_{max}$ due to magnetic
ordering of {\it cerium} and normalized coefficient of $-\log T$
term, $m$($P$)/$m$(0.1) originated from Kondo-like scattering of
conduction electrons decrease with increasing pressure. Further,
the $T_{\rm max}$ disappear at a critical pressure ($P_{\rm c}$
$\sim$ 0.9 GPa) where there is a corresponding change in the slope
of the coefficient of the $-\log T$ term, indicating a decrease in
the strength of the Kondo-like scattering. The suppression of
$T_{\rm max}$ has been attributed to a valence change of {\it
cerium} from Ce$^{3+}$ state to Ce$^{4+}$/Ce$^{3+}$ mixed valence
state under pressure.

\end{document}